# Silicon-based metasurfaces for aperture-robust spectrometer/imaging with angle integration

Weizhu Xu, Qingbin Fan, Peicheng Lin, Jiarong Wang, Hao Hu, Tao Yue ,

Xuemei Hu, and Ting Xu

**Abstract**

Compared with conventional grating-based spectrometers, reconstructive spectrometers based on spectrally engineered filtering have the advantage of miniaturization because of the less demand for dispersive optics and free propagation space. However, available reconstructive spectrometers fail to balance the performance on operational bandwidth, spectral diversity and angular stability. In this work, we proposed a compact silicon metasurfaces based spectrometer/camera. After angle integration, the spectral response of the system is robust to angle/aperture within a wide working bandwidth from 400nm to 800nm. It is experimentally demonstrated that the proposed method could maintain the spectral consistency from F/1.8 to F/4 (The corresponding angle of incident light ranges from ±7° to ±16°) and the incident hyperspectral signal could be accurately reconstructed with a fidelity exceeding 99%. Additionally, a spectral imaging system with 400×400 pixels is also established in this work. The accurate reconstructed hyperspectral image indicates that the proposed aperture-robust spectrometer has the potential to be extended as a high-resolution broadband hyperspectral camera.

**Introduction**

Optical spectrometer is one of the most important instruments for materials analysis because of the abundant information contained in the absorption spectrum of materials. It was widely applied in food quality control [1], environmental monitoring [2], and art conservation [3]. Conventional spectrometers are mostly established on dispersive optics or Fourier transform. For the former, dispersive optics such as dispersive prisms or gratings are utilized to disperse the incident light to the sensor plane, and then the intensity of corresponding frequency components is directly detected by photoelectric sensors. The latter modulates the incident light on a single detector over time with an interferometer and transforms the received signals to a wavelength-dependent spectrum via Fourier transform. Conventional spectrometers usually rely on the combination of dispersive optics, detector arrays, and movable parts, which limited their applications because of the demand for free propagation space and consumption on time.

Over the past decade, a new spectrometer paradigm relies on computational techniques has emerged. This type of spectrometer does not measure the frequency components directly, instead it utilizes a set of broadband filters to encode the spectral information, and then reconstructs the original spectral signal by solving a linear equation. This type of reconstructive spectrometer does not require dispersive optics or mechanical components and could be easily integrated into the imaging sensor for miniaturization. The key point of this paradigm is how to design a set of appropriate spectral filters. Many optical elements that can generate diverse spectral response functions could be adopted for spectrometry systems, such as nanowires [4-5], quantum dots [6-7], photonic crystal slabs [8] liquid crystals [9] and metasurfaces [10]. Particularly in the field of metasurfaces, the subwavelength nanoscale units have an exceptional capacity to manipulate the phase, amplitude, polarization, or spectrum of incident electromagnetic waves [11-12]. However, due to the limit on the characteristics of materials and the two-dimension structure optimization strategy, these filters [13-15] often fail to balance the operational bandwidth, spectral diversity and angular stability. Searching for the perfect filter array is a challenging task.

Past works [13-15] focus on the optimization of complex two-dimensional structures. In calibration, they assume that the incident light comes from a planar light source. As the angle of incident light changes, the response of the metasurfaces based filters also changes significantly. Simulated results shown in Fig. 1 demonstrate that there is a set of irregular peaks in absorption spectrum of the these filters, which will be shifted as the incident angle changes. Meanwhile, we found that in a spectrometer/camera, the incident light is focused to the sensor plane by a set of optical lens and it is no longer parallel after passing through these apertures. In this case, the robustness on convergence angle of the incident light, or the aperture size is the main variable of the system, which is significantly different from the assumptions of existing work. Especially in spectral imaging, where the object is fixed relative to the sensor, aperture stability is more significant than angle stability. As shown in Fig.1b, the input light from a solid angle is focused on the filter array and the angle-sensitive spike signals in Fig. 1e will be integrated as an aperture-robust spectral response or smoothed out. The simulated results on silicon metasurfaces based filter in Fig. 1f illustrate that changing the aperture of the system has little effect on the spectral response. Based on the simulation we further propose a compact spectrometer containing 256 C-silicon metasurfaces based filters which are composed of periodically arranged nanoscale pillars of different sizes. Complex structural optimizations such as topology-optimization [19] are not required here. Experiments also demonstrate that the spectral response is stable when aperture ranges from F/4 to F/1.8 (The corresponding angle of incident light ranges from ±7° to ±16°) which is a common aperture size in photography. The spectral consistency at different aperture is larger than 99.79%. We also make another larger metasurfaces array with a size of 400x400 pixels for hyperspectral imaging. The proposed system works well in spectrum ranging from 400 nm to 800 nm, which balances the performance on operational bandwidth, spectral diversity and angular/aperture stability, and simplifies the design of nanostructures. Focusing lenses for angle integration are compatible with current spectrometers/cameras and additional components are not required.

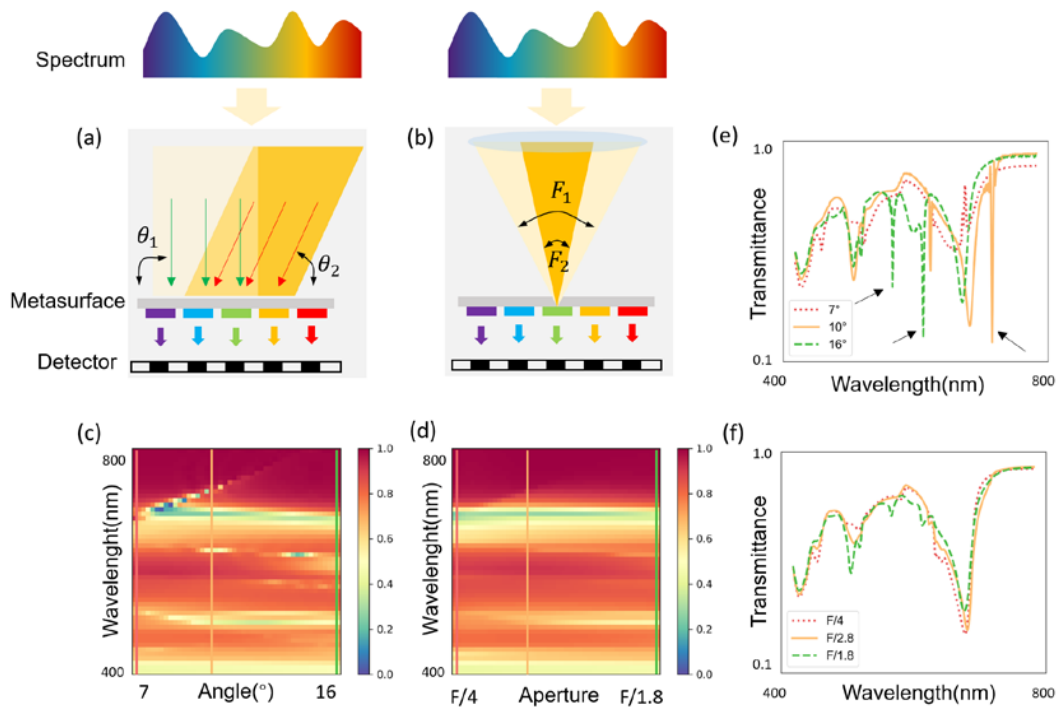

**Figure 1 | The transmittance of the C-silicon metasurfaces based filter on two kind of incident light.** (a) Planar light with different incident angle. (b) Focused light with different numerical aperture. The metasurfaces based filter is composed of periodically arranged C-silicon nanopillars. The spectral transmittance at different angles and apertures are shown in (c) and (d). (e) shows the spectral transmittance curves at three specific incident angles, 7°, 10°, and 16°, and the subfigure (f) shows those at the corresponding apertures F/4, F/2.8, F/1.8. After the angle integration, the absorption peaks denoted by black arrows in (e) are smoothed out in (f), thus enhancing the aperture stability of the system.

## Operation scheme

The operational principle of the spectrometer includes spectral coding and computational reconstruction. The former is illustrated in Fig. 1(a-b), where the incident light is transmitted sequentially through the optical lens, the metasurfaces based filters and the sensor. When the spectral transmittance of the filter does not depend on the incident angle, this process could be described by the following equation:

$$\begin{aligned}I_1 &= \int_1^m L(\lambda)M_1(\lambda)D(\lambda)S(\lambda)d\lambda \\ I_2 &= \int_1^m L(\lambda)M_2(\lambda)D(\lambda)S(\lambda)d\lambda \\ &\vdots \\ I_n &= \int_1^m L(\lambda)M_n(\lambda)D(\lambda)S(\lambda)d\lambda\end{aligned} \quad \text{(Eq. 1)}$$

$S(\lambda)$ represents the spectrum of incident light. $L(\lambda)$ denotes the transmittance of the optical components in the system, including imaging lens or optical filters. $M(\lambda)$ represents the transmittance of the metasurfaces based filters and $D(\lambda)$ is the quantum efficiency of the sensor. Eq.1 can be organized as a typical linear equation:

$$I = AS, A \in \mathbb{R}^{n,m} \quad \text{(Eq. 2)}$$

$A$ is the system transmission matrix, which is obtained by calibration. $n$ is the number of metasurfaces based filters and $m$ is the number of spectral channels. In our experiments, n and m are set to 100 and 400 respectively. When n>m, Eq.2 is overdetermined and could be solved by least squares methods [20]. When n<m, Eq.2 degrades to an underdetermined equation and additional assumptions on $S$, such as smoothing constraints are required here.

Sparse priori in the gradient domain is a common constrain to solve the underdetermined equation:

$$\hat{S} = (A^T A + \eta J)^{-1} A^T I \quad \text{(Eq. 3)}$$

$$J = \begin{bmatrix} 1 & -1 & 0 & \cdots & 0 \\ 0 & 1 & -1 & \ddots & \vdots \\ 0 & 0 & \ddots & \ddots & 0 \\ \vdots & \ddots & 0 & 1 & -1 \\ 0 & \cdots & 0 & 0 & 1 \end{bmatrix}$$

Here $J$ is the gradient matrix which constrains the gradient sparsity on reconstructed spectrum. $\eta$ is the weight on gradient constrain. The larger $\eta$, the smoother the results. Eq.3 is the weighted least square solution of Eq.2. It is computationally fast to solve but not accurate enough in spectral reconstruction. Here we introduce the sparse optimization and dictionary learning [21-22] method to improve the results. In dictionary learning [23], it is assumed that the input signal is linearly composed of a set of bases learned from a large spectrum dataset and the weight on these bases is sparse. The linear equations can be converted to the $l_1$-norm minimization problem:

$$\hat{S} = \operatorname{argmin} \|A\Phi\psi - I\|_2^2 + \eta\Gamma(\psi). \quad \text{Eq.4}$$

where $\Phi$ is a sparse representation matrix and $\psi$ is the weight of the input spectrum on $\Phi$. The symbol "$\Gamma$" denotes the $l_1$-norm minimization. In the experiment, the sparse representation matrix $\Phi$ is consisted of a set of Gaussian shaped curves with a center wavelength from 350nm to 820 nm and a bandwidth from 2nm to 200nm. We used an open-source package for solving the convex optimization problem [24-25].

When the spectral transmittance of the filter is angle sensitive, $M(\lambda)$ should be rewritten as $M(\lambda, \theta)$. After angle integration, the transmittance curve of the paraxial filters can be represented as:

$$M'(\lambda, \theta') = \frac{2}{\tan^2\theta'} \int_0^{\theta'} \frac{\sin\theta}{\cos^3\theta} M(\lambda, \theta) \, d\theta \quad \text{(Eq. 5)}$$

$\theta'$ is the incident angle related to the maximum aperture. In addition, we define the spectral consistency within a certain range of angles/apertures as the minimum value of the correlation coefficient of the normalized transmittance:

$$\mu(M) = \min_{1 \leq i \neq j \leq n} |\langle M_i, M_j \rangle| \quad (Eq.\,6)$$

$i$ and $j$ denote two incident angles or apertures. To evaluate the performance of spectral reconstruction, we also introduce a metric of reconstruction fidelity defined as:

$$F(S, \hat{S}) = \langle S, \hat{S} \rangle \quad (Eq.\,7)$$

where $S, \hat{S}$ are the input and reconstructed spectral signal after normalization and <> denotes the inner product.

**Experiments**

**1. Simulation and fabrication.** To explore the aperture dependence of silicon-based metasurfaces, we employ the Finite-Difference Time-Domain (FDTD) method to simulate the transmittance of periodic nanopillars. Fig. 1(c) depicts the spectral transmittance distribution of the nanopillar with a diameter of 130 nm, a height of 500nm and a period of P=400 nm. C-silicon based nanopillars exhibit low intrinsic losses and support multiple strong resonant modes within the entire visible spectrum, resulting in high transmittance and multiple distinctive peaks in the device's spectral performance. According to Eq. 5, we could calculate the transmittance distribution on aperture after angle integration. As shown in Fig. 1b, the input light from a solid angle is focused on the filter array and the angle-sensitive spike signals will be integrated as an aperture robust spectral response. These simulated results demonstrate that the transmittance from F/4 to F/1.8 remain consistent and the spectral consistency $\mu(M)$ on aperture is 99.65%, larger than the corresponding consistency on angle, which is only 96.72%.

The designed metasurfaces unit cell is composed of an array of monocrystalline silicon nanopillars with a height of 500 nm, situated on an alumina substrate, arranged in a square lattice. We have designed a total of 256 different metasurfaces filters, with each metasurfaces filter containing two distinct sizes of nanostructures. To generate a diverse spectral response function, the dimensions of these nanostructures vary between 80 and 300 nm, with a periodicity range of 330 to 450 nm. The fabrication of the silicon-based metasurfaces proposed here primarily involves the utilization of electron beam lithography (EBL), inductively coupled plasma (ICP) etching, and thin film deposition techniques. Fig. 2 shows microscopic images of the fabricated metasurfaces device. In Fig. 2(a), an optical microscopy image displays the pixelated metasurfaces broadband filters. Figs. 2(b) and 2(c) are scanning electron microscopy images that showcase well-defined nano-pillar structures.

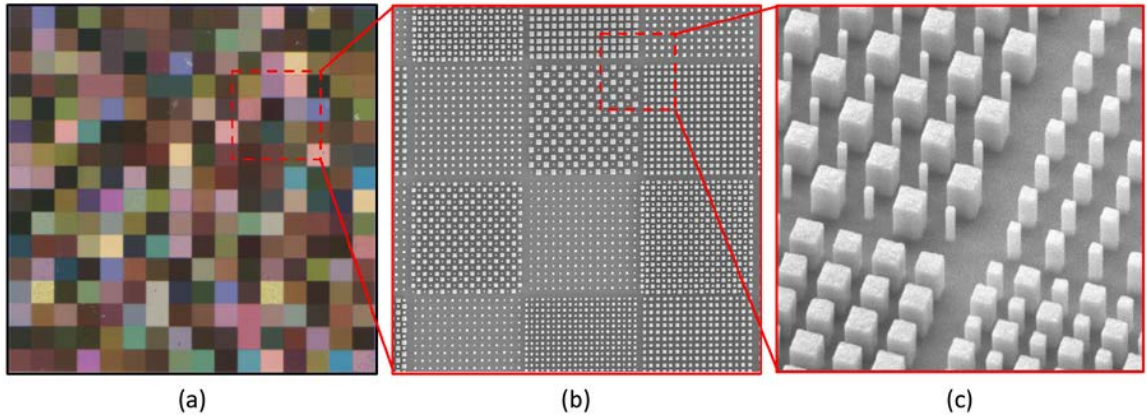

**Figure 2 | Microscopic images of the fabricated silicon-based metasurfaces device.** (a) Optical microscopy image of the pixelated metasurfaces device. (b and c) Scanning electron microscopy images depicting the top view and oblique view of the nanopillars.

## 2.Calibration

We utilized the Xenon lamp and monochromator to calibrate the spectral transmission matrix **A** of the system. The prototype is shown in Fig. 3. Firstly, the broadband light source from the Xenon lamp is converted to uniform narrowband light by the monochromator and the integrating sphere. After passing through the main lens and the meta-filters, the signal is captured by an image sensor (Grasshopper3 GS3-U3-123S6M). A commercial spectrometer (Ideaoptics PG2000-Pro) is also equipped on the integrating sphere to record the intensity information of the input light source.

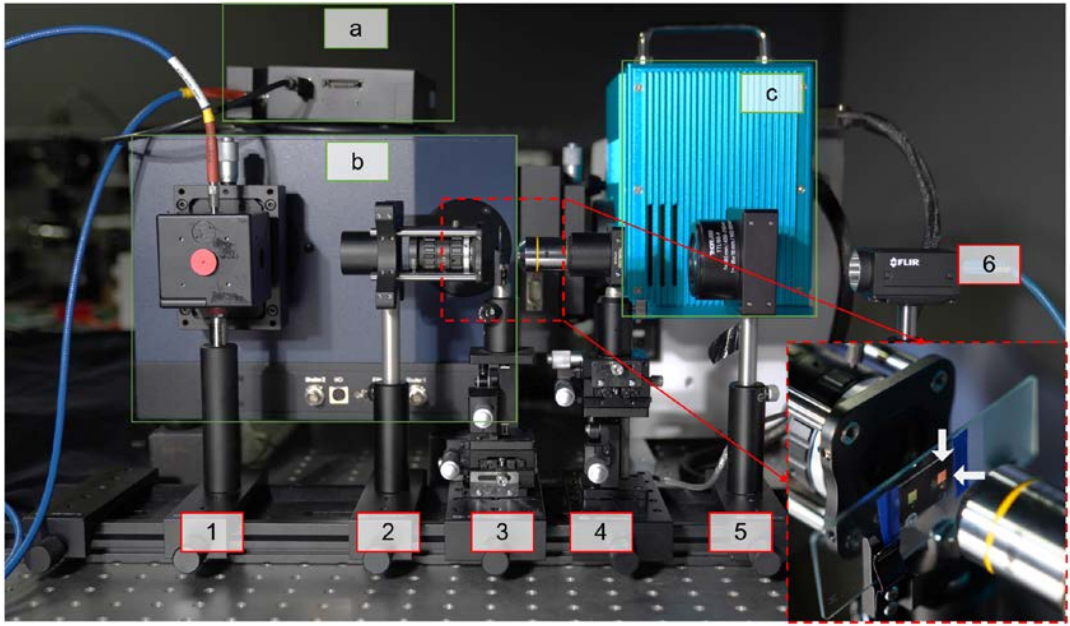

**Figure 3 | The optical setup of the proposed spectrometer/imaging system.** (a-c) in the green box is the narrowband light source for calibration, containing (a) a commercial spectrometer, (b) a monochromator and (c) a Xenon lamp. Below are the main components, containing (1) an integrating sphere, (2) main lens, (3) C-silicon metasurfaces based filter array, (4) microscope lens, (5) tube lens, and (6) an image sensor. Currently, the meta-filter array is not integrated into the sensor and relay lens are utilized here. The red dotted box is a zoomed-in view where the proposed meta-filter array is located between the white arrows. The integrating sphere is only used for system calibration. It will be removed in the spectral imaging experiments.

The spectral transmittance curves of 16 randomly selected structures under two different apertures (F/4, F/1.8) are shown in Fig. 4 and the quantum efficiency of the sensor has been taken into account here. The average transmittance of the system is around 33% within the spectrum from 400nm to 800nm. Compared with the narrow-band filter, our system has higher signal-to-noise ratio (SNR), which is significant for high-precision spectrum reconstruction. To the best of our knowledge, the system has the widest working bandwidth in visible spectrometer based on metasurfaces. For the aperture stability, it can be observed that the transmittance distributions from F/4 to F/1.8 remain consistent. The average spectral consistency $\mu(A)$ of the 256 kinds of meta-structures is as high as 99.79%. For the rest of the article, we utilize the spectral transmittance at F/1.8 as the system transmission matrix $A$. Additionally, we define the spectral diversity as the L2 norm of the correlation matrix between the normalized spectral curves:

$$\sigma(A) = \frac{1}{n}\|AA^T - E_n\|_2 \text{ (Eq. 8)}$$

$E_n$ is the n-dimensional identity matrix. After calibration, the spectral diversity of our system is 0.83 when n is set to 100. The lower the spectral diversity, the higher the stability on the solution of the linear equation Eq.2. The correlation matrix at n=100 is shown in Fig. 5.

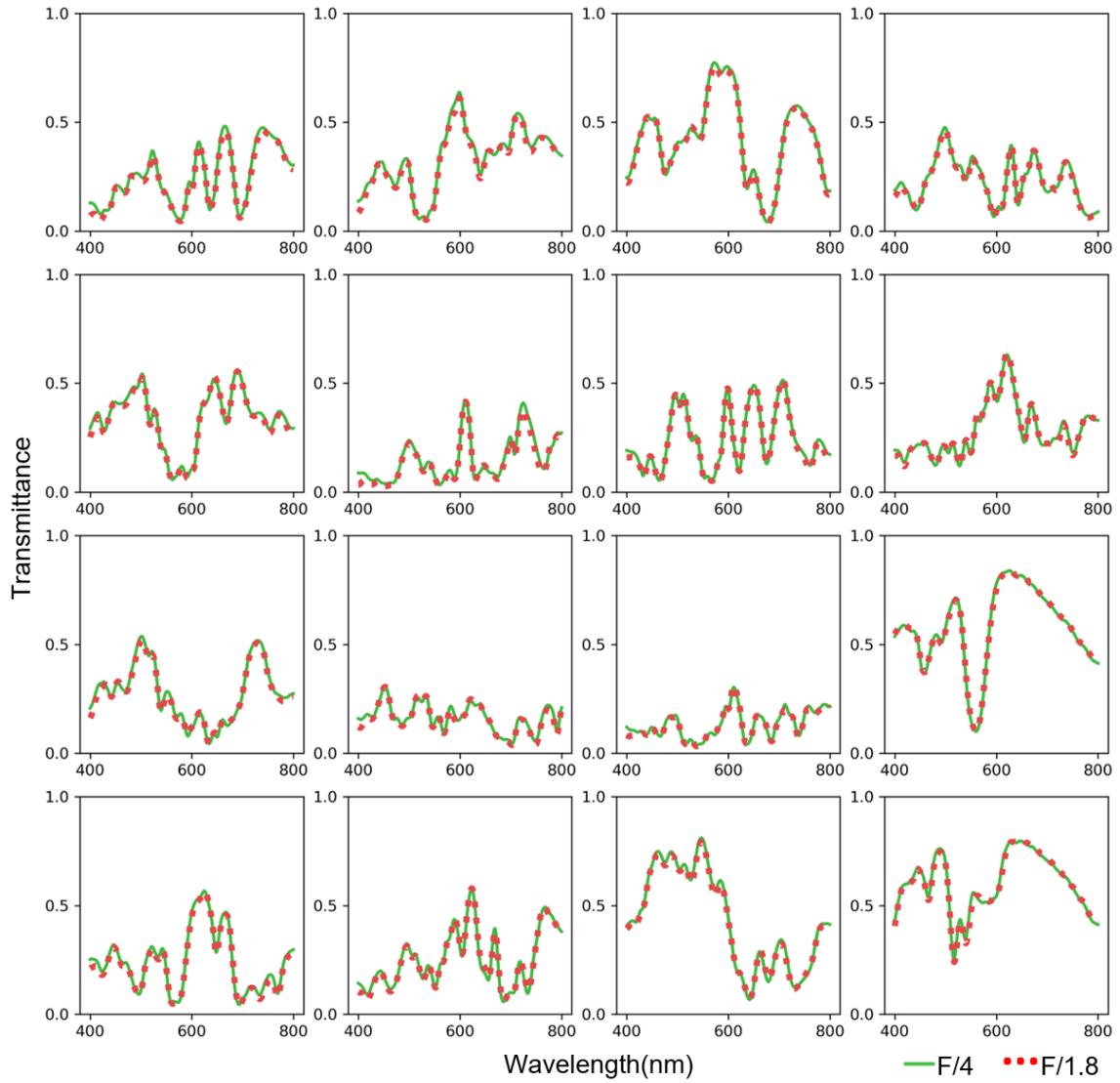

**Figure 4 | The transmittance curves of the proposed spectrometer with aperture of F/4 (the green line) and F/1.8 (the red dotted line).** The high aperture stability of our system across a broad bandwidth from 400nm to 800 nm ensures the high reconstruction accuracy.

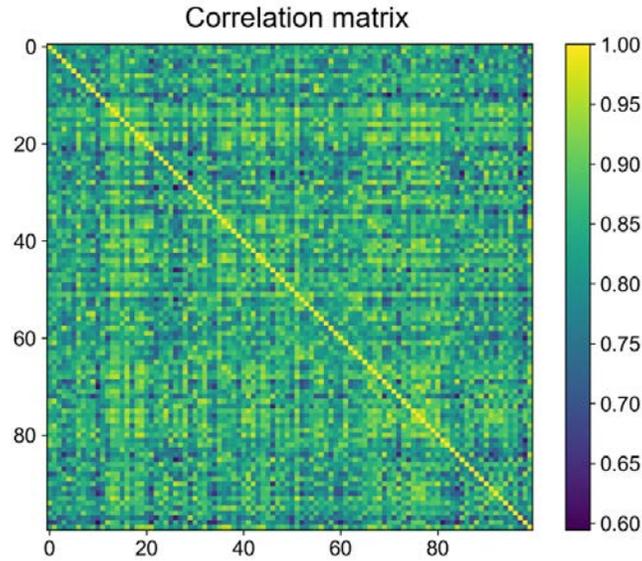

**Figure 5 | The correlation matrix of the normalized spectral curves on 100 selected structures.**

### 3. Performance on spectral reconstruction

Data preprocessing is required before spectral reconstruction. The broken area and structures with low transmittance are ignored. Similar structures with high correlation coefficients are also excluded. The spectral response curves of the selected high-quality filters are shown in Fig. 6. It could be found that from n=36 to n=100, the spectral response of the selected structures distributed within the wavelength from 400nm to 800nm uniformly and reaches a similar spectral diversity about 0.82. The 256 kind of structures can be flexibly selected according to the demand for spectral resolution.

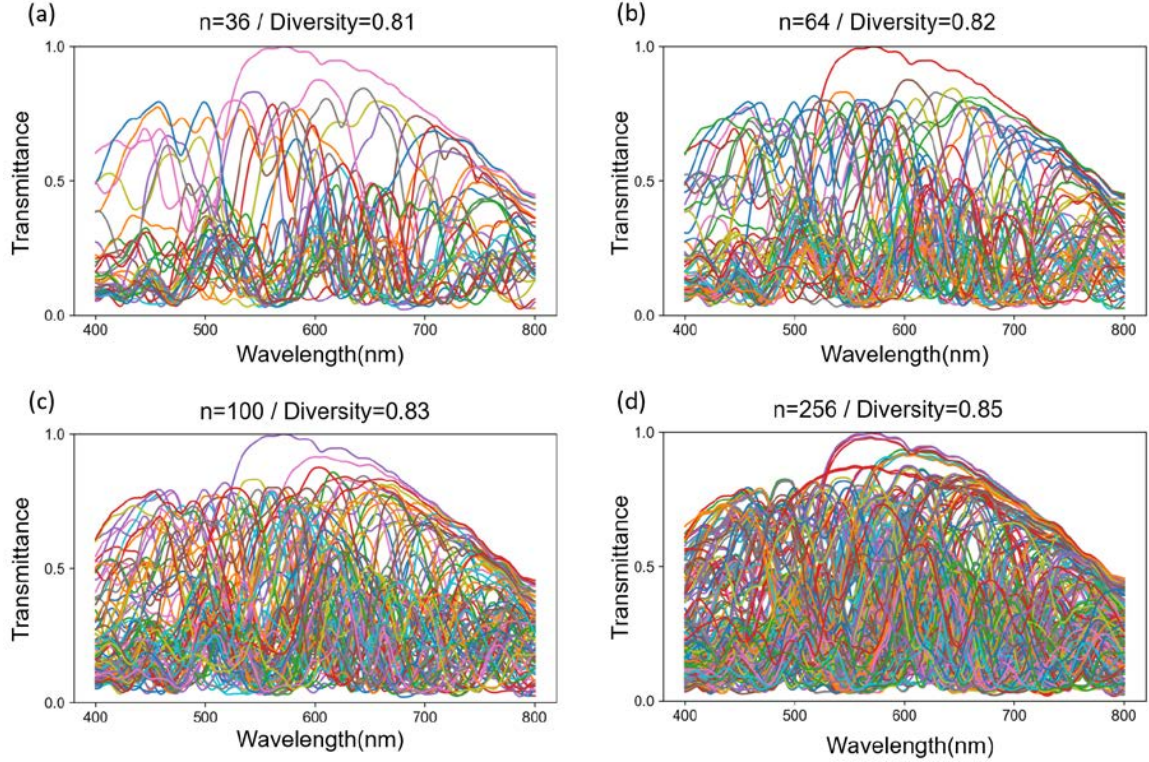

**Figure 6 | The spectral response on different number of selected structures.** The corresponding spectral diversity is also labeled in figure titles.

In our experiments, 100 kinds of structures in Fig. 6(c) are selected for hyperspectral reconstruction. Firstly, we test the performance on two light emitting diodes (LED): a broad-spectrum white LED and a band-pass green LED. We calibrate the spectrum of the light sources as ground truth with a commercial spectrometer and captured the signals at F/1.8 and F/4. In order to verify the extreme performance of the system, we additionally capture the signals at the aperture of F/8, which corresponds to an incident angle of ±3.6°. The dictionary learning method in Eq. 4 is adopted as the reconstruction algorithm. The reconstructed spectral curves are shown in Fig. 7 and the fidelity of the reconstructed signals is shown in Table 2. These results demonstrated that our system still has a high reconstruction accuracy even at the aperture of F/8.

**Table 2 | Fidelity of the reconstructed signals on three apertures.**

| Fidelity | F/1.8 | F/4 | F/8 |
|---|---|---|---|
| **White LED** | 99.55% | 99.52% | 98.88% |
| **Green LED** | 99.65% | 99.49% | 99.42% |

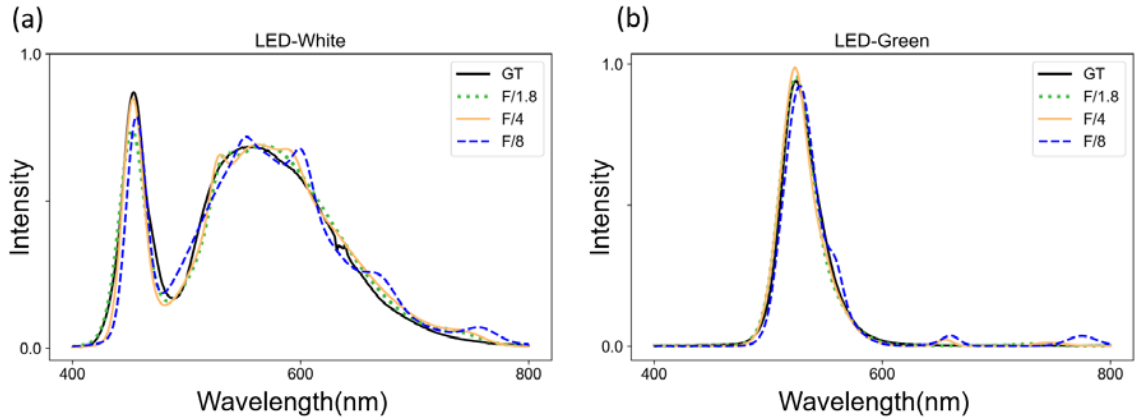

**Figure 7 | The reconstructed spectrum on white and green LEDs.** The black line denotes the ground truth calibrated by a commercial spectrometer.

### 3. Performance on spectral Imaging

Another meta-filter array with a size of 400x400 pixels is also fabricated for hyperspectral imaging. The size of each filter is about 6.9um*6.9 um, smaller than that in spectral measurement which is 35um x35um. The target scene in Fig. 8a is a colorful plate containing 24 color blocks, and the aperture is set to F/1.8. The spectral transmission matrix of the array is calibrated with the same method mentioned in **Sec. 2**. The hyperspectral image shown in Fig. 9 is reconstructed by the dictionary learning method point by point. In order to reconstruct these points at the edges of the color blocks, we introduce the edge detection method adopted in [14] to select neighboring regions adaptively for spectral reconstruction. In the experiment each region contains about 40-50 structures.

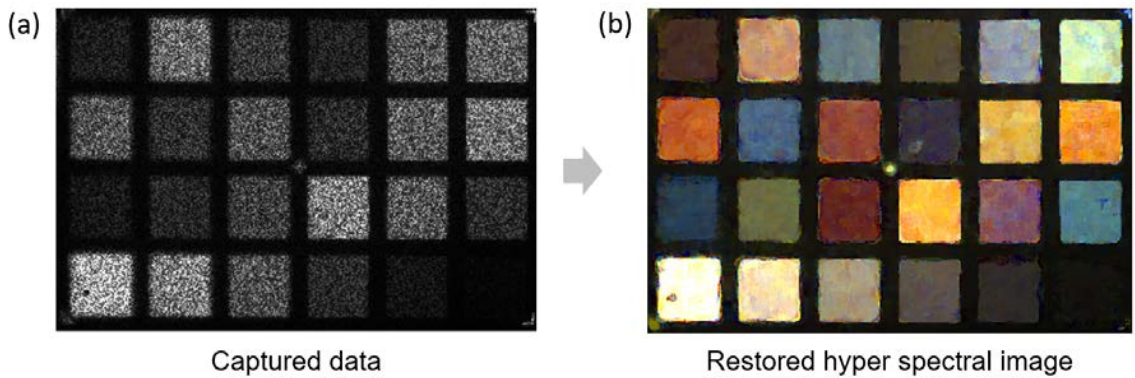

**Figure 8 | Spectral reconstruction on a colorful plate.** The coded signal captured by the camera is shown on the left, and the reconstructed pseudo-color image is shown on the right.

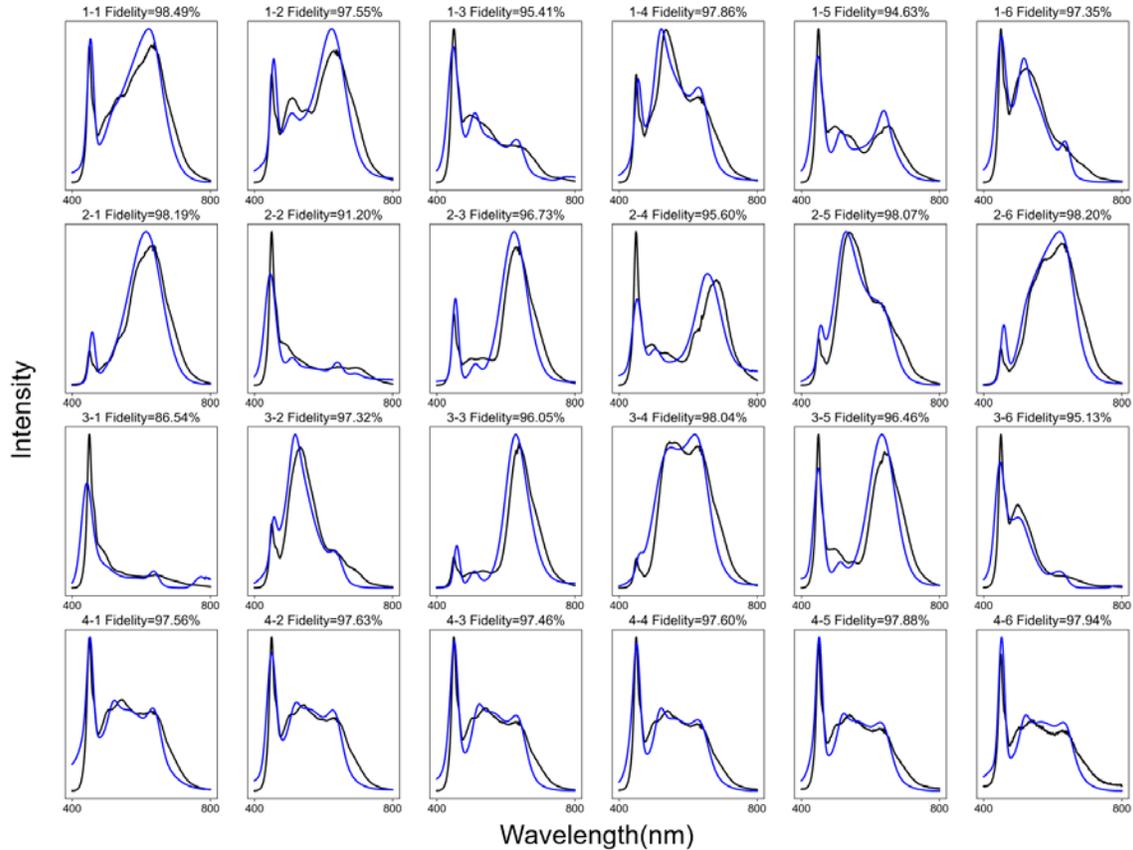

**Figure 9 | Spectral reconstruction results at the center of the 24 color blocks.** The black and blue curves in the figure are the ground truth and the reconstruction result. The fidelity of each image is labeled in the titiles of the corresponding subimage.

It could be found that the accuracy on spectral imaging is a little worse than the spectral measurement because of the smaller pixel size in the 400x400 filter-array. Specifically, there are crosstalks between smaller pixels due to aberrations in the optical system and the signal-to-noise ratios (SNR) of the imaging system are also lower than that in spectral measurement. Additionally, the number of pixels that we select for reconstruction in the spectral imaging is about 40-50, less than that in spectral measurement which contains 100 structures, and this also affects the reconstruction accuracy. Nevertheless, our system still achieves an average reconstruction fidelity about 96.45% and has the potential to be a high-resolution hyperspectral camera.

**Conclusion**

The proposed C-silicon metasurfaces based spectrometer/camera works well on a wide operational bandwidth from 400nm to 800nm and performs excellent on spectral diversity. In this work, it is also demonstrated that the spectral response of these filters has a high aperture stability after angle integration because angle sensitive absorption peaks will be integrated as aperture robust responses or smoothed out. In spectral imaging system, the position of the optics and sensor is fixed and the only variable affecting the system transmission matrix is the aperture rather than the incident angle. Aperture stability is significant in photograph. The average consistency of the system transmission matrix from F/1.8 to F/4 is as high as 99.79%, and experiments have demonstrated that even at larger aperture F/8, the average reconstruction fidelity is still up to 98.88%. Compared with the existing metasurfaces based spectral filters, the proposed system does not rely on complex structural optimization strategies and could still support the high transmittance, the wide working bandwidth and the weak spectral correlation, which results in high signal-to-noise ratio and numerical stability. In conclusion, we propose a concise spectrometer/camera working on a wide operational bandwidth from 400nm to 800nm exhibiting a high systematic stability. The flexible and compact system framework also shows great potential for this C-silicon metasurfaces based filters to be extended as a single shot high-resolution hyperspectral camera.

**Acknowledgments**

The work is supported by the Key Research and Development Program from Ministry of Science and Technology of China (2022YFA1205000, 2022YFA1207200), National Natural Science Foundation of China (11774163, 61971465, 61671236, 12104225), Natural Science Foundation of Jiangsu Province of China (BK20220068, BK20212004, BK20200304) and Fundamental Research Funds for the Central Universities.